\documentstyle[12pt,epsfig]{article}
\newcommand{\be}[1]{\begin{equation} \label{(#1)}}
\newcommand{\ee}{\end{equation}}
\newcommand{\ba}[1]{\begin{eqnarray} \label{(#1)}}
\newcommand{\ea}{\end{eqnarray}}
\newcommand{\nn}{\nonumber}

\def\lsim{\mbox{${}^< \hspace*{-7pt} _\sim$}}

\def\Lfv{$L_f\hspace{-0.95em}/\ \ $}

\def\lg{\langle}
\def\rg{\rangle}
\def\m{$\mu^--e^-$}
\begin{document}

\begin{center}
{\Large\bf Nuclear $\mu^--e^-$ conversion in strange quark sea}\\

\bigskip

{T.S. Kosmas$^a$, Sergey Kovalenko$^{b}$\footnote{On
leave of absence from the Joint Institute for Nuclear Research, Dubna, Russia}
and Ivan Schmidt$^b$} \\[0.5cm]
{$^a$\it Division of Theoretical Physics, University of Ioannina GR-45110
Ioannina, Greece}\\
{$^b$\it Departamento de F\'\i sica, Universidad
T\'ecnica Federico Santa Mar\'\i a, Casilla 110-V, Valpara\'\i so, Chile}
\end{center}

\bigskip

\begin{abstract}
We study nuclear $\mu^--e^-$ conversion in the general framework 
of effective Lagrangian approach without referring to any specific realization 
of the physics beyond the standard model (SM) responsible for lepton flavor 
violation (\Lfv). All the possible types of short range interactions 
(non-photonic mechanisms), i.e. (pseudo-)scalar, (axial-)vector  
and tensor, are included in our formalism.    
We show that the \m conversion in the strange nucleon sea via the scalar interactions  
is comparable with that in the valence  quarks. This provides an insight into the strange 
quark couplings beyond the SM. From the available experimental data on $\mu^--e^-$ conversion 
and expected sensitivities of planned experiments we derived upper bounds on the generic 
\Lfv - parameters of \m conversion sensitive to the relevant u-,d- and s-quark couplings.
\end{abstract}

\bigskip
\bigskip


\bigskip
\bigskip


\newpage
%%%%%%%%%%%%%%%%%%%%%%%%%%%%%%%%%%%%%%%%%%%%%%%%%%%%%%%%%%%%%%%%%%%%%%%%%%%%

The muon-flavor violating processes 
\begin{equation}
\mu^- + (A,Z) \longrightarrow  e^- \,+\,(A,Z)^*\, ,
\label{I.1}
\end{equation}
i.e. muon-to-electron ($\mu^--e^-$) conversion in nuclei,
is known as a very sensitive probe of lepton flavor violation (\Lfv) 
and related physics beyond the standard model (SM) \cite{LFV}-\cite{FKKV:00}. 
This fact has been recently strengthened by the
evidence for the muon-neutrino oscillations, 
drawn by the Superkamiokande experiment, 
which is the first convincing signal of the non-standard physics 
connected to the lepton flavor non-conservation. 
The distinct feature of coherent enhancement in nuclear \m 
conversion makes it more promising probe of \Lfv than other 
lepton flavor violating processes 
($\mu^-\rightarrow e^-\gamma$, etc.).
In general the structure of a participating nucleus    
brings some uncertainties into the theoretical predictions 
for \m conversion. 
However, in the most interesting case of coherent \m conversion 
\cite{mu-e-nucl} these uncertainties can be significantly reduced by the 
possibility of using the available experimental data on nuclear densities. 

On the experimental side, at present, there is one running \m conversion 
experiment, SINDRUM II \cite{SINDRUM}, and two planned experiments, MECO 
\cite{Molzon,MECO} and PRIME \cite{PRIME}. The SINDRUM II experiment at PSI 
\cite{SINDRUM} with $^{48}$Ti as stopping target has established the 
best upper bound on the branching ratio 
\begin{eqnarray}\label{Ti}
&&R_{\mu e}^{Ti} = \frac{\Gamma(\mu^- + {}^{48}Ti\rightarrow e^- + {}^{48}Ti)}
{\Gamma(\mu^- + {}^{48}Ti
\rightarrow \nu_{\mu} + ^{48}Sc)} \leq 6.1\times 10^{-13}\ , \ \ \ 
\mbox{(90\% C.L.)} \ \ \  \cite{SINDRUM}.
\end{eqnarray}
The MECO experiment with $^{27}$Al is going to start soon at Brookhaven \cite{MECO}.
The sensitivity of this experiment is expected to reach the limit 
\begin{eqnarray}\label{Al}
R_{\mu e}^{Al}  = \frac{\Gamma(\mu^- + {}^{27}Al\rightarrow e^- +  {}^{27}Al)}{\Gamma(\mu^- + {}^{27}Al
\rightarrow \nu_{\mu} +  {}^{27}Mg)} \leq 2\times 10^{-17}\ \ \ \ \ \ \ \   \cite{MECO}
\end{eqnarray}
This year the PSI experiment is running with the very heavy nucleus $^{197}$Au
aiming to improve by a factor of about 20-30 over the previous limit, 
$R_{\mu e}^{Au} \leq 2.0 \times 10^{-11}$, set on \m  in $^{197}$Au 
by the same experiment some years ago \cite{SINDRUM,Vintz}. 
Now the expected limit is  
\begin{eqnarray}\label{Au} 
R_{\mu e}^{Au} = \frac{\Gamma(\mu^- + ^{197}Au\rightarrow e^- 
+ ^{197}Au)}{\Gamma(\mu^- + ^{197}Au \rightarrow 
\nu_{\mu} + ^{197}Pt)} \leq 6 \times 10^{-13}\ \ \ \ \ \ \  \cite{SINDRUM,Vintz}
\end{eqnarray}
Very recently, a proposal for a new experiment at Tokyo (PRIME) was announced 
\cite{PRIME}. It intends to utilize the $^{48}$Ti as stopping target with the
impressive expected sensitivity of $R_{\mu e}^{Ti} \leq  10^{-18}$ \cite{PRIME}.

These experimental limits can put severe constraints on mechanisms of \m 
conversion.  In the literature there have been studied various mechanisms 
beyond the SM (see \cite{LFV}-\cite{mu-e-nucl} and references therein)
classified into two categories: photonic and non-photonic as shown in Fig. 1. 
Specific mechanisms from both categories significantly
differ in many respects and, in particular, in
nucleon and nuclear structure treatment. This is attributed to
the fact that they operate at different
distances and, therefore, involve different details of the nucleon structure.
Long-distance photonic mechanisms (Fig. 1(a)) are mediated by virtual
photon exchange between the nucleus and the $\mu-e$ lepton current.
They suggest that the \m conversion occurs in the lepton-flavor non-diagonal
electromagnetic vertex which is presumably induced by non-standard model physics 
at loop level. The hadronic vertex is characterized in this case by ordinary 
electromagnetic nuclear form factors.
Contributions to $\mu-e$ conversion via virtual
photon exchange exist in all models which allow $\mu\rightarrow e \gamma$
decay.
On the other hand, short-distance non-photonic mechanisms (Fig. 1(b)) are 
described by the effective \Lfv $ $ 4-fermion quark-lepton interactions 
which may appear after integrating 
out heavy intermediate states ($W,Z$, Higgs bosons, supersymmetric particles etc.). 

In this Letter we focus on the non-photonic mechanisms of \m conversion 
(Fig. 1(b)). The generic effect of physics beyond the SM in \m 
is described in our approach by an effective Lagrangian which includes 
all the possible 4-fermion quark-lepton interactions. 
Our special interest is concentrated on the scalar interactions which
are sensitive to the heavy quark content of the nucleon. 
Also, the current \m experiments at Brookhaven and PSI can efficiently probe 
the scalar component of the \m conversion \cite{SINDRUM}-\cite{Vintz}. 
We will show that the contribution to the \m conversion rate which originates 
from the strange nucleon sea via the scalar interactions is comparable 
with that coming from the valence quarks of the nucleon.   

We derive a general representation of the \m branching ratio in terms of 
generic \Lfv $ $ parameters of the effective 4-fermion quark-lepton \m transition 
operators. Transforming these operators, first to the nucleon and then to the 
nuclear level, we pay special attention to the effects of nucleon and nuclear 
structure. 
The nucleon structure is taken into account on the basis of the QCD 
picture of baryon masses and experimental data on certain hadronic parameters. 
For nuclear structure calculations we apply the formalism described in 
Refs. \cite{tsk-mon}-\cite{mu-e-nucl}. 
Our applications refer to the nuclei of current experimental interest, $^{27}$Al, 
$^{48}$Ti and $^{197}$Au, with special attention to $^{197}$Au  which has not been previously
studied in the context of \m conversion.
\begin{figure}[h!]
\vspace{-2 cm}
\hspace{-0.5 cm}
\mbox{ \epsfxsize=16 cm\epsffile{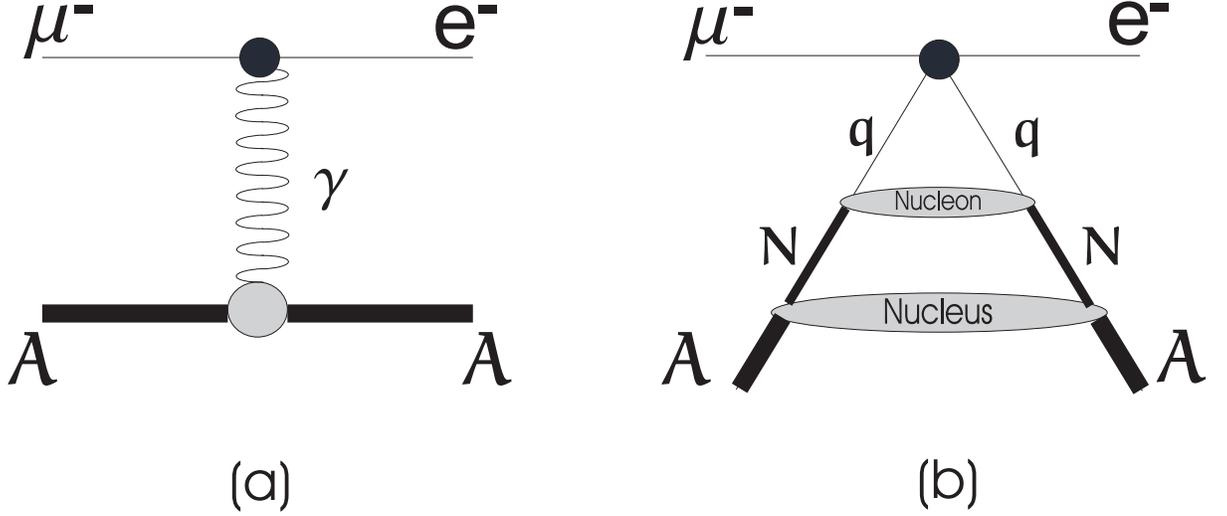}}
\vspace{-0.5cm}
\caption{ (a) Photonic(long-distance) and (b) non-photonic(short-distance) 
contributions to the nuclear $\mu^-- e^-$ conversion.}
\end{figure}
We start with the 4-fermion effective Lagrangian describing 
the non-photonic $\mu-e$ conversion at the quark level which corresponds to the upper 
vertex of the diagram in Fig. 1(b). It can be written in a general Lorentz covariant form as
\begin{eqnarray}
{\cal L}_{eff}^{q}\ =\  \frac{G_F}{\sqrt{2}}  
\sum_{A,B,C,D;q}\ \left[ \eta_{AB}^{(q)} j_{\mu}^A\ J_{(q)}^{B\mu} + 
 \eta_{AB}^{(q)} j^A\ J_{(q)}^{B} +
 \eta_{T}^{(q)} j_{\mu\nu}\ J_{(q)}^{\mu\nu}\right].
\label{eff-q}
\end{eqnarray}  
where the summation involves $A,B = \{A,V\}$, $C,D =\{S,P\}$ and $q= \{u,d,s\}$.   
The \Lfv parameters $\eta_i^q$ depend on a concrete \Lfv model.
The lepton and quark currents are
$
%\begin{eqnarray}
j_{\mu}^V = \bar e \gamma_{\mu} \mu, \ 
j_{\mu}^A = \bar e \gamma_{\mu}\gamma_5 \mu,  \ 
j^S = \bar e \ \mu, \ 
j^P = \bar e \gamma_{5} \mu, \ 
j_{\mu\nu}^V = \bar e \sigma_{\mu\nu} \mu,\ $ 
%\nn
%  
$J_{(q)}^{V\mu} = \bar q \gamma^{\mu} q, \  
J_{(q)}^{A\mu} = \bar q \gamma^{\mu}\gamma_5 q, \  
J_{(q)}^{S} = \bar q \ q, \ 
J_{(q)}^{P} = \bar q \gamma_5\ q, \ 
J_{(q)}^{\mu\nu} = \bar q \sigma^{\mu\nu} q.
%
%\label{currents-q}
%\end{eqnarray}
%
$

The next step is the reformulation of the quark level 
Lagrangian (\ref{eff-q}) in terms of the nucleon effective fields. 
First we write down the nucleon level Lagrangian in a general Lorentz covariant form
with the isospin structure of the \m transition operator
\begin{eqnarray} \nn
{\cal L}_{eff}^{N}\ &=&\   \frac{G_F}{\sqrt{2}}
\sum_{A,B,C,D}\  \left[j_{\mu}^A (\alpha_{AB}^{(0)} J_{(0)}^{B\mu} + 
\alpha_{AB}^{(3)} J_{(3)}^{B\mu}) + 
j^C (\alpha_{CD}^{(0)} J_{(0)}^{D} + 
\alpha_{CD}^{(3)} J_{(3)}^{D}) +\right.\\  
\label{eff-N}  
&+& \left. j_{\mu\nu} (\alpha_{T}^{(0)} J_{(0)}^{\mu\nu} + 
\alpha_{T}^{(3)} J_{(3)}^{\mu\nu})\right].
\end{eqnarray}
The isoscalar  $J_{(0)}$ and isovector $J_{(3)}$ nucleon currents are defined as 
$
%\begin{eqnarray}\nn
J_{(k)}^{V\mu} = \bar N \gamma^{\mu}\tau_k N, \ \  
J_{(k)}^{A\mu} = \bar N \gamma^{\mu}\gamma_5 \tau_k N,\ \ 
J_{(k)}^{S} = \bar N \tau_k N, \ \ 
J_{(k)}^{P} = \bar N \gamma_5 \tau_k N,\ \ 
J_{(k)}^{\mu\nu} = \bar N \sigma^{\mu\nu}\tau_k N, 
%
%\label{Nucl-curr}
%\end{eqnarray}
$
where $ k = 0,3$ and $\tau_0 \equiv \hat I$. 
In Eq. (\ref{eff-N}) we neglected derivative terms. In the matrix 
elements of $\mu^--e^-$ conversion they produce a small contribution 
proportional to $q/m_p\leq m_{\mu}/m_p \sim 0.1$ where $q=|{\bf q}|$ is 
the momentum transfer to the nucleon while $m_{\mu}$ and $m_p$ are the muon and 
the proton masses respectively. 
 
Now we relate the coefficients $\alpha$ in Eq. (\ref{eff-N}) with the ``fundamental"
\Lfv parameters $\eta$ of the quark level Lagrangian (\ref{eff-q}).
Towards this end we apply the on-mass-shell matching condition \cite{FKSS97}
\begin{equation}
\langle \Psi_F|{\cal L}_{eff}^{q}|\Psi_I\rangle \approx \langle \Psi_F|{\cal L}%
_{eff}^{N}|\Psi_I\rangle ,  
\label{match}
\end{equation}
where $|\Psi_I\rangle $ and $\langle \Psi_F|$ are the initial and
final nucleon states.

In order to solve this equation we use
various relations for the matrix elements of the quark operators
between the nucleon states
\begin{eqnarray}\label{mat-el1}
\langle N|\bar{q}\ \Gamma_{K}\ q|N\rangle = G_{K}^{(q,N)}
\bar{\Psi}_N\ \Gamma_{K}\ \Psi_N,
\end{eqnarray}
with $q=\{u,d,s\}$,  $N=\{p,n\}$ and  $K = \{V,A,S,P,T\}$,
$\Gamma_K = \{\gamma_{\mu}, \gamma_{\mu}\gamma_5, 1, \gamma_5, \sigma_{\mu\nu}\}$.
Since the maximum momentum transfer
in $\mu -e$ conversion is much smaller than the typical scale of 
nucleon structure we can safely neglect the ${\bf q}^{2}$-dependence of 
the nucleon form factors $G_{K}^{(q,N)}$ and drop the weak magnetism as well as 
the induced pseudoscalar terms proportional to the small momentum transfer.

Isospin symmetry requires that
\begin{eqnarray}\label{isosym}
G_{K}^{(u,n)}=G_{K}^{(d,p)}\equiv G_{K}^{d}, \ \ \
G_{K}^{(d,n)}=G_{K}^{(u,p)}\equiv G_{K}^{u},\ \ \
G_{K}^{(s,n)}=G_{K}^{(s,p)}\equiv G_{K}^{s}.
\end{eqnarray}
Note that the axial, pseudoscalar and tensor nucleon currents couple to the nuclear spin  
leading, therefore, to incoherent contributions. Thus, only the vector and 
scalar nucleon form factors are needed in the case of the coherent nuclear $\mu^--e^-$ 
conversion which we are studying in the present paper.

Conservation of vector current implies that the vector charge is equal to the number of 
the valence quarks of the nucleon and, therefore, 
\begin{eqnarray}\label{cvc}
G_{V}^{u}=2,\ \ \  G_{V}^{d}=1, \ \ \ G_{V}^{s}=0.
\end{eqnarray}

The scalar form factors can be extracted from the baryon octet $B$ mass
spectrum $M_B$ in combination with the data on the pion-nucleon sigma term
%\begin{eqnarray}\label{sigma}
$
\sigma_{\pi N} = (1/2)(m_u + m_d)\lg p|\bar u u + \bar d d|p\rg.
%\end{eqnarray}
$
%Towards this end 
%In this way 
We follow the QCD picture of the baryon masses 
based on the relation 
$
%\begin{eqnarray}\label{emt}
\lg B|\theta^{\mu}_{\mu}|B\rg = M_B \bar B B
%\end{eqnarray}
$
%\cite{SVZ,Cheng}
and on the well known representation \cite{SVZ} for the trace of
the energy-momentum tensor 
$
%\begin{eqnarray}\label{trace}
\theta^{\mu}_{\mu} = m_u \bar u u + m_d \bar d d
+ m_s \bar s s - (\tilde b\alpha_s/8\pi)G_{\mu\nu}^a G^{\mu\nu}_a,
%\end{eqnarray}
$
where $G_{\mu\nu}^a$ is the gluon field strength, $\alpha_s$ is
the QCD coupling constant and $\tilde b$ is the reduced Gell-Mann-Low
function with the heavy quark contribution subtracted. Using these
relations in combination with $SU(3)$ relations \cite{Cheng} for
the matrix elements $\lg B|\theta^{\mu}_{\mu}|B\rg$  as well as
the experimental data on $M_B$ and  $\sigma_{\pi N}$ we derive
\begin{eqnarray}\label{scalar}
G_S^{u} \approx 5.1, \ \ \ G_S^{d} \approx 4.3, \ \ \  G_S^{s} \approx 2.5
\end{eqnarray}
for the conventional values of the current quark masses:
$m_u= 4.2$ MeV, $m_d = 7.5$ MeV, $m_s = 150$ MeV. These masses, however,
are not yet well determined \cite{PDG}  and uncertainty factor $\lsim$ 2 may affect the
results in Eq. (\ref{scalar}). For our further numerical estimates these uncertainties 
are not critical and we take the central values Eq. (\ref{scalar}) for the 
form factors $G_S^{u,d,s}$. 
The approach of Refs. \cite{Nath}, relying on the results of the lattice simulations, 
gives for these form factors similar values with the same level of uncertainties.
What remains true is that the strange quarks of the nucleon sea significantly contribute 
to the scalar nucleon form factor $G_S$. This result dramatically differs from
the na\"{\i}ve quark model and the MIT bag model where $G_{A,S,P}^{s} = 0$.

Now the above Eq. (\ref{match}) can be solved and the coefficients $\alpha$ of the   
nucleon level Lagrangian (\ref{eff-N}) can be expressed in terms of the generic \Lfv   
parameters $\eta$ of the quark level effective Lagrangian Eq. (\ref{eff-q}) as         
\begin{eqnarray}
\label{alpha}
\alpha_{IV}^{(0)} &=&\frac{1}{2}(\eta_{IV}^{(u)} + \eta_{IV}^{(d)}) (G_{V}^{u} + G_{V}^{d}),
\ \ \ \ \ \ \  
\alpha_{IV}^{(3)} = \frac{1}{2}(\eta_{IV}^{(u)} - \eta_{IV}^{(d)}) (G_{V}^{u} - G_{V}^{d}),
\\ \nn
\alpha _{JS}^{(0)} &=& \frac{1}{2}(\eta_{JS}^{(u)} + \eta_{JS}^{(d)}) (G_{S}^{u} + G_{S}^{d})
+ \eta_{JS}^{(s)} G_{S}^{s},
\ \ \
\alpha _{JS}^{(3)} = \frac{1}{2}(\eta_{JS}^{(u)} - \eta_{JS}^{(d)}) (G_{S}^{u} - G_{S}^{d}),
\end{eqnarray}
where $I=V,A$ and $J=S,P$.

From the Lagrangian (\ref{eff-N}), following the standard procedure,
we can derive the formula for the total $\mu-e$ conversion branching ratio. In this paper
we confine ourselves to the coherent process which is the dominant channel of 
$\mu-e$ conversion exhausting, for the majority of experimentally interesting nuclei,
more than $90\%$ of the total \m  branching ratio \cite{mu-e-nucl}.
To leading order of the non-relativistic reduction the coherent
$\mu-e$ conversion branching ratio takes the form
\begin{equation}
R_{\mu e^-}^{coh} \ = \ 
\frac{G_F^2} {4 \pi } \ {\cal Q} \  
\frac{p_e E_e \ ({\cal M}_p + {\cal M}_n)^2 }
{ \Gamma ({\mu^-\to capture}) }
\, ,
\label{Rme}
\end{equation}
where $p_e, E_e$ are the outgoing electron 3-momentum and energy. We define 
the quantity 
\begin{eqnarray}
{\cal Q} \,=\, |\alpha_{VV}^{(0)}+\alpha_{VV}^{(3)}\ \phi|^2 +
|\alpha_{AV}^{(0)}+\alpha_{AV}^{(3)}\ \phi|^2 +
|\alpha_{SS}^{(0)}+\alpha_{SS}^{(3)}\ \phi|^2 + 
|\alpha_{PS}^{(0)} + \alpha_{PS}^{(3)}\ \phi|^2 
\nonumber  \\
+ 2\ {\rm Re}\{(\alpha_{VV}^{(0)}+\alpha_{VV}^{(3)}\ \phi)(\alpha_{SS}^{(0)}+
\alpha_{SS}^{(3)}\ \phi)^*
+  (\alpha_{AV}^{(0)}+\alpha_{AV}^{(3)}\ \phi)(\alpha_{PS}^{(0)}+\alpha_{PS}^{(3)}\ \phi)^*\}\, .
\label{Rme.1}
\end{eqnarray}
The nuclear transition matrix elements ${{\cal M}}_{p,n}$
entering Eq. (\ref{Rme}), for the case of a ground state to ground 
state \m transition, are defined as 
\begin{equation}
{\cal M}_{p,n} = 4\pi \int j_0(p_e r) \Phi_\mu (r) \rho_{p,n} (r) r^2  dr \ ,
\label{V.1}
\end{equation}
($j_0$ is the zero-order spherical Bessel function) where
$\rho_{p,n}(r)$ are the spherically symmetric proton (p) and neutron (n) 
nuclear densities normalized to the atomic number $Z$ and neutron number 
$N$, respectively, of the participating nucleus. 
$\Phi_{\mu}(r)$ is the space dependent part of the muon wave function.

The quantity ${\cal Q}$ in (\ref{Rme.1}) depends on the nuclear-target
parameters through the factor 
% \begin{eqnarray}\label{phi}
$\phi = ({\cal M}_p - {\cal M}_n)/({\cal M}_p + {\cal M}_n) \, .$
% \approx (Z-N)/(Z+N),
% \end{eqnarray}
However for all the experimentally interesting nuclei this parameter is small. Therefore
nuclear dependence of ${\cal Q}$ can always be neglected except very special
narrow domain in the \Lfv parameter space where $\alpha^{(0)}\leq \alpha^{(3)}\phi$.
Another important issue of the smallness of the ratio $\phi$ in Eq. (\ref{Rme.1})
is the dominance of the isoscalar contribution associated with the coefficients 
$\alpha^{(0)}$. Given that the relative significance of the strange quark component, 
which enters only the isoscalar couplings, is enhanced.

The nuclear matrix elements ${\cal M}_{p,n}$, defined in Eq. (\ref{V.1}), 
are numerically calculated using proton densities $\rho_{p}$ from Ref.
\cite{Vries} and neutron densities $\rho_{n}$ from Ref. \cite{Gibbs}
whenever possible. 
The muon wave function $\Phi_\mu(r)$ was obtained by 
solving the Schr\"ondinger equation with the Coulomb potential produced by the densities 
$\rho_{p,n}$, taking into account the vacuum polarization corrections \cite{FKKV:00}.  
In this way the matrix elements ${\cal M}_{p,n}$ for the nuclear targets $^{27}$Al and $^{48}$Ti
have been calculated in Ref. \cite{FKKV:00}. Here we apply this approach to $^{197}$Au. 
The results for ${\cal M}_{p,n}$ corresponding to the 
nuclei Al, Ti and Au are given in \mbox{Table 1} where
we also show the muon binding energy $\epsilon_b$ and the experimental 
total rates ($\Gamma_{\mu c}$) of the ordinary muon capture reaction \cite{Suzuki}.

By inserting in Eq. (\ref{Rme}) the values of the nuclear matrix elements 
${\cal M}_{p,n}$ and other quantities from Table 1 and Eqs. (\ref{Ti})-(\ref{Au})
we can derive the upper limits on the parameters of the effective 
Lagrangians (\ref{eff-q}), (\ref{eff-N}). These limits correspond to
the existing \cite{SINDRUM} or expected \cite{MECO,PRIME} experimental 
limits on the branching ratio $R_{\mu e}$.
The most straightforward limit can be set on the quantity ${\cal Q}$ of 
Eq. (\ref{Rme}) but its physical meaning is rather obscure. In order to obtain
physically interesting upper bounds on the \Lfv parameters $\alpha^{(0)}$, 
$\eta^{(q)}$ we adopt the usual assumption that different terms in expressions 
(\ref{alpha}) and (\ref{Rme.1}) do 
not substantially compensate each other or, equivalently, that only one term 
dominates at a time.  In this way we derived the upper limits given in Table 2.
The scaling factors $B_{A}$ in Table 2 are defined as
\begin{eqnarray}
\label{scale}
B_{Ti} = \left(\frac{R^{exp}_{\mu e}}{6.1 \cdot 10^{-13}}\right)^{1/2},\ \ 
B_{Al} = \left(\frac{R^{exp}_{\mu e}}{2.0 \cdot 10^{-17}}\right)^{1/2}, \ \ 
B_{Au} = \left(\frac{R^{exp}_{\mu e}}{6.0 \cdot 10^{-13}}\right)^{1/2}.
\end{eqnarray}
Multiplying corresponding column in Table 2 by $B_{A}$ one can reconstruct upper 
limits on the listed parameters for the case when experimental upper bounds on the 
branching ratio $R^{exp}_{\mu e}$ differ from those we used in our analysis.

The constraints for $\eta_{JS}^{(s)} $ in Table 2 originate from the contribution 
of the strange nucleon sea. As can be seen, they are comparable 
to the other \m limits derived from the valence quarks contributions.

The limits in Table 2 represent a general outcome of the $\mu^--e^-$ conversion experiments 
for the \Lfv physics. These limits can be easily translated into limits on the parameters 
of specific \Lfv model predicting the $\mu^--e^-$ conversion. This is achieved by 
adjusting the quark level effective Lagrangian of the model to the form of Eq. (\ref{eff-q}) 
and by identifying the effective parameters $\eta_{AB}^{(q)}$ with expressions in terms of 
model parameters. Then the upper bounds on $\eta_{AB}^{(q)}$ from Table 2 can 
be translated to certain constraints on the model parameters present in these expressions. 

In conclusion, we constructed a general  
effective Lagrangian describing the non-photonic $\mu^--e^-$ conversion and specified 
all the \Lfv parameters characterizing this process without referring 
to any specific model beyond the SM. It includes (axial-)vector, (pseudo-)scalar 
and tensor interactions. We derived general formula for the coherent 
$\mu^--e^-$ conversion branching ratio in terms of the \Lfv 
parameters of the quark level effective Lagrangian.
We calculated previously unknown nuclear matrix elements of  
$^{197}$Au (the current SINDRUM II target). We found that the $\mu^--e^-$ 
conversion branching ratio in the strange quark sea of the nucleon is comparable 
with that in the valence quarks. We have inferred the generic $\mu^--e^-$ constraints on 
the \Lfv parameters from the existing and expected experimental 
bounds on the $\mu^--e^-$ conversion rate. These results 
are readily used for derivation of the constraints on 
the parameters of any specific \Lfv model without need of the nucleon and nuclear 
structure calculations. This provides useful nuclear and 
particle physics inputs for the expected new data from the PSI, MECO, PRIME 
and other $\mu^--e^-$ experiments to set bounds on the muon \Lfv violating parameters.
 
\bigskip
This work was supported in part by Fondecyt (Chile) under grant
8000017, by a C\'atedra Presidencial (Chile) and by RFBR (Russia) under 
grant 00-02-17587. T.S.K. would like to express his appreciation to UTFSM  
for hospitality.

%%%%%%%%%%%%%%%%%%%%%%%%%%%%%%%%%%%%%%%%%%%%%%%%%%%%%%%%%%%%%%%%%%%%%%%%%%%%%%%

%%%%%%%%%%%%%%%%%%%%%%%%%%%%%%%%%%%%%%%%%%%%%%%%%%%%%%%%%%%%
% 

\newpage
%%%%%%%%%%%%%%%%%%%%%%%%%%%%%%%%%%%%%%%%%%%%%%%%%%%%%%%%%%%%%
\begin{table}%[ht]
\begin{center}
\begin{tabular}{rccccc}
\hline \hline
   &  &  &  &  &  \\
Nucleus & $p_e \, (fm^{-1})$ & $\epsilon_b \, (MeV)$ &
$\Gamma_{\mu c} \, ( \times 10^{6} \, s^{-1})$ &
${\cal M}_p$ & ${\cal M}_n$  \\
\hline
   &  &  &  &  &  \\
$^{27}$Al  & 0.531 &  -0.470 &  0.71 & 0.047 & 0.045   \\
   &  &  &  &  &  \\
$^{48}$Ti  & 0.529 &  -1.264 &  2.60 & 0.104 & 0.127   \\
   &  &  &  &  &  \\
$^{197}$Au & 0.485&   -9.938 & 13.07 & 0.395 & 0.516   \\
\hline
\hline
\end{tabular}
\caption{Transition nuclear matrix elements ${\cal M}_{p,n}$ (in $fm^{-3/2}$)
of Eq. (\ref{V.1}) 
and other useful quantities
(see the text).}
\end{center}
\end{table}
%%%%%%%%%%%%%%%%%%%%%%%%%%%%%%%%%%%%%%%%%%%%%%%%%%%%%%%%%%%%%%%%%%%%%%

\vskip -3cm                                                                   
 
%%%%%%%%%%%%%%%%%%%%%%%%%%%%%%%%%%%%%%%%%%%%%%%%%%%%%%%%%%%%%%%%%%%%%% 
\begin{table} 
\begin{center} 
\begin{tabular}{|c|c|c|c|} 
\hline \hline 
Parameter & Present limits&Expected limits  & Expected limits \\ 
&$^{48} $Ti$(\mu-e)\cdot B_{Ti}$  & $^{27}$Al$(\mu-e)\cdot B_{Al}$  
&$^{197}$Au$(\mu-e)\cdot B_{Au}$ \\ 
\hline 
&& &    \\ 
${\cal Q}$ & $2.1\times 10^{-14}$&$1.2\times 10^{-18}$&$7.8 \times 10^{-15}$\\  
&& &    \\ 
$\alpha_{_{AB}}^{(0)}$&$ 1.5\times 10^{-7}$&$1.1\times 10^{-9}$&$8.5\times 10^{-8}$\\ 
&& &    \\ 
$\eta_{_{IV}}^{(u,d)}$&$ 1.0 \times 10^{-7}$&$7.3 \times 10^{-10}$&$ 5.7\times 10^{-8}$\\ 
&& &    \\ 
$\eta_{_{JS}}^{(u,d)}$&$ 3.0 \times 10^{-8}$&$2.3 \times 10^{-10}$&$ 1.8 \times 10^{-8}$\\ 
&& &    \\ 
$\eta_{_{JS}}^{(s)}  $&$ 5.8 \times 10^{-8}$&$4.4 \times 10^{-10}$&$ 3.4\times 10^{-8}$\\ 
&& &    \\ 
\hline         
\hline
\end{tabular}
\caption{Upper bounds on the \Lfv parameters (see definitions in the text) 
inferred from the SINDRUM II data on $^{48}$Ti [Eq. (\ref{Ti})] 
as well as from the expected sensitivities of the current SINDRUM II 
[Eq. (\ref{Au})] and future MECO  [Eq. (\ref{Al})] experiments employing
$^{197}$Au and $^{27}$Al as stopping targets respectively. We denoted: $AB = VV,AV,SS,PS$; 
$I=V,A$; $J=S,P$. The scaling factors $B_{i}$ are defined in Eq. (\ref{scale}).}
\end{center}
\end{table}
%%%%%%%%%%%%%%%%%%%%%%%%%%%%%%%%%%%%%%%%%%%%%%%%%%%%%%%%%%%%%%%%%%%%%%

\end{document}